\documentclass[11pt]{article}

\usepackage[final]{acl}

\usepackage{times}
\usepackage{latexsym}

\usepackage[T1]{fontenc}

\usepackage[utf8]{inputenc}

\usepackage{microtype}

\usepackage{inconsolata}

\usepackage{graphicx}

\usepackage{booktabs}
\usepackage{graphicx}
\usepackage{amsmath}
\usepackage{tikz}
\usepackage{pgfplots}
\definecolor{myblue}{RGB}{31,119,180}
\definecolor{myorange}{RGB}{255,127,14}
\definecolor{mygreen}{RGB}{44,160,44}
\definecolor{myred}{RGB}{214,39,40}
\definecolor{mypurple}{RGB}{148,103,189}
\definecolor{mybrown}{RGB}{140,86,75}
\definecolor{mypink}{RGB}{227,119,194}
\definecolor{mygray}{RGB}{127,127,127}
\definecolor{myolive}{RGB}{188,189,34}
\definecolor{mycyan}{RGB}{23,190,207}
\newcommand{\modelname}{CleanCodec}
\newcommand{\B}{\textbf}
\newcommand{\U}{\underline}

\title{\modelname: Efficient and Robust Speech Tokenization via Perceptually Guided Encoding}

\author{Eugene Kwek \\
  Pennsylvania State University \\
  \texttt{eugene.kwek.1@gmail.com} \\ \And
  Feng Liu \\
  Drexel University \\
  \texttt{fl397@drexel.edu} \\ \AND
  Rui Zhang \\
  Pennsylvania State University \\
  \texttt{rmz5227@psu.edu} \\ \And
  Wenpeng Yin \\
  Pennsylvania State University \\
  \texttt{wenpeng@psu.edu} \\}

\begin{document}
\maketitle
\begin{abstract}
Neural audio codecs are a key component of speech processing pipelines, compressing audio into discrete tokens for downstream modeling. However, existing codecs struggle to balance reconstruction quality with token efficiency, often encoding perceptually irrelevant information such as background noise and recording artifacts at the expense of linguistically and acoustically meaningful content. We reframe audio tokenization as a selective information bottleneck problem and propose \modelname, a \textit{denoising} audio codec which learns to encode only perceptually important features and discard imperceptible information. At just 12.5 tokens per second, \modelname~achieves state-of-the-art tokenization efficiency, substantially outperforming existing codecs in speaker similarity and speech intelligibility. Evaluations on downstream text-to-speech and voice conversion tasks further demonstrate improved performance and up to 17x faster inference, highlighting significant efficiency gains.\footnote{Inference and training code will be released on Github.}. 
\end{abstract}

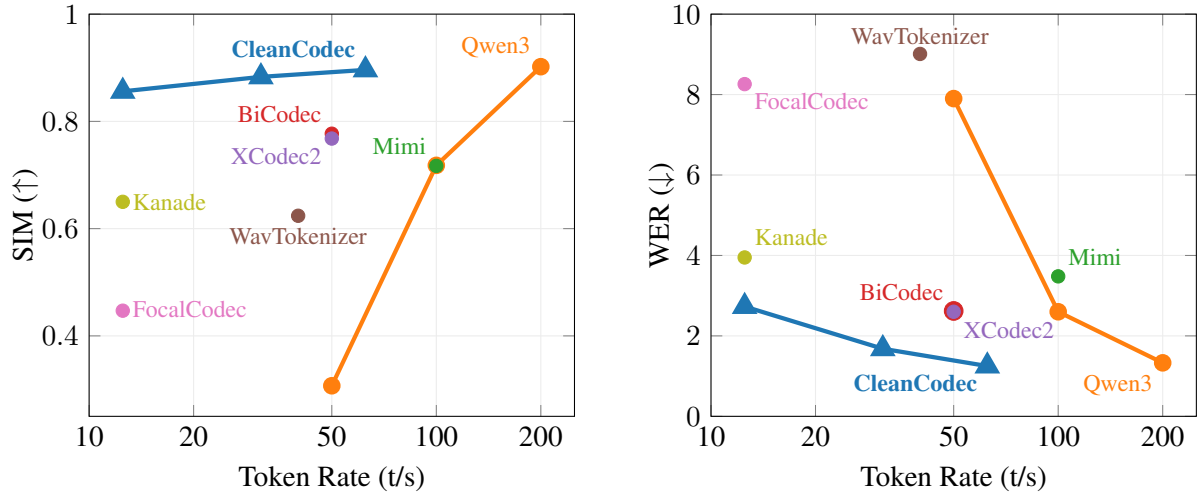
\begin{figure*}[h]
\begin{center}

\begin{minipage}{0.48\linewidth}
\centering

\begin{tikzpicture}
\begin{axis}[
    width=8cm,
    xmode=log,
    xlabel={Token Rate (t/s)},
    ylabel style={yshift=-1em},
    ylabel={SIM ($\uparrow$)},
    xmin=10, xmax=250,
    ymin=0.25, ymax=1.00,
    xtick={10, 20, 50, 100, 200},
    xticklabels={10, 20, 50, 100, 200},
    grid=both,
    grid style={line width=0.2pt, draw=gray!30},
    major grid style={line width=0.4pt, draw=gray!15},
    clip=false,
]
 
\addplot[
    color=myblue, ultra thick,
    mark=triangle*, mark size=4pt,
    mark options={fill=myblue},
] coordinates {
    (12.5,  0.856)
    (31.25, 0.883)
    (62.5,  0.896)
};
\node[myblue, anchor=south east, font=\small] at (axis cs:62.5,  0.90) {\textbf{\modelname}};
 
\addplot[
    color=myorange, ultra thick,
    mark=*, mark size=2.5pt,
    mark options={fill=myorange},
] coordinates {
    (50,  0.307)
    (100, 0.718)
    (200, 0.902)
};
\node[myorange, anchor=south east, font=\small] at (axis cs:200, 0.90) {Qwen3};
 
\addplot[
    color=mygreen, mark=*, mark size=2.5pt,
    mark options={fill=mygreen}, only marks,
] coordinates {(100, 0.717)};
\node[mygreen, anchor=south east, font=\small] at (axis cs:100, 0.72) {Mimi};
 
\addplot[
    color=myred, mark=*, mark size=2.5pt,
    mark options={fill=myred}, only marks,
] coordinates {(50, 0.777)};
\node[myred, anchor=south east, font=\small] at (axis cs:50, 0.78) {BiCodec};
 
\addplot[
    color=mypurple, mark=*, mark size=2.5pt,
    mark options={fill=mypurple}, only marks,
] coordinates {(50, 0.768)};
\node[mypurple, anchor=north east, font=\small] at (axis cs:50, 0.77) {XCodec2};
 
\addplot[
    color=mybrown, mark=*, mark size=2.5pt,
    mark options={fill=mybrown}, only marks,
] coordinates {(40, 0.624)};
\node[mybrown, anchor=north, font=\small] at (axis cs:40, 0.62) {WavTokenizer};
 
\addplot[
    color=mypink, mark=*, mark size=2.5pt,
    mark options={fill=mypink}, only marks,
] coordinates {(12.5, 0.447)};
\node[mypink, anchor=west, font=\small] at (axis cs:12.5, 0.45) {FocalCodec};
 
\addplot[
    color=myolive, mark=*, mark size=2.5pt,
    mark options={fill=myolive}, only marks,
] coordinates {(12.5, 0.650)};
\node[myolive, anchor=west, font=\small] at (axis cs:12.5, 0.65) {Kanade};
 
\end{axis}
\end{tikzpicture}
\label{fig:sim}
\end{minipage}
\hfill
\begin{minipage}{0.48\linewidth}
\centering

\begin{tikzpicture}
\begin{axis}[
    width=8cm,
    xmode=log,
    xlabel={Token Rate (t/s)},
    ylabel={WER ($\downarrow$)},
    ylabel style={yshift=-1.5em},
    xmin=10, xmax=250,
    ymin=0, ymax=10,
    xtick={10, 20, 50, 100, 200},
    xticklabels={10, 20, 50, 100, 200},
    grid=both,
    grid style={line width=0.2pt, draw=gray!30},
    major grid style={line width=0.4pt, draw=gray!15},
    clip=false,
]
 
\addplot[
    color=myblue, ultra thick,
    mark=triangle*, mark size=4pt,
    mark options={fill=myblue},
] coordinates {
    (12.5,  2.73)
    (31.25, 1.68)
    (62.5,  1.25)
};
\node[myblue, anchor=north east, font=\small] at (axis cs:62.5,  1.3)  {\textbf{\modelname}};
 
\addplot[
    color=myorange, ultra thick,
    mark=*, mark size=2.5pt,
    mark options={fill=myorange},
] coordinates {
    (50,  7.90)
    (100, 2.60)
    (200, 1.33)
};
\node[myorange, anchor=north east, font=\small] at (axis cs:200, 1.3) {Qwen3};
 
\addplot[
    color=mygreen, mark=*, mark size=2.5pt,
    mark options={fill=mygreen}, only marks,
] coordinates {(100, 3.48)};
\node[mygreen, anchor=south west, font=\small] at (axis cs:100, 3.5) {Mimi};
 
\addplot[
    color=myred, mark=*, mark size=3.5pt,
    mark options={fill=myred}, only marks,
] coordinates {(50, 2.62)};
\node[myred, anchor=south east, font=\small] at (axis cs:50, 2.62) {BiCodec};
 
\addplot[
    color=mypurple, mark=*, mark size=2.5pt,
    mark options={fill=mypurple}, only marks,
] coordinates {(50, 2.60)};
\node[mypurple, anchor=north west, font=\small] at (axis cs:50, 2.6) {XCodec2};
 
\addplot[
    color=mybrown, mark=*, mark size=2.5pt,
    mark options={fill=mybrown}, only marks,
] coordinates {(40, 9.01)};
\node[mybrown, anchor=south, font=\small] at (axis cs:40, 9.0) {WavTokenizer};
 
\addplot[
    color=mypink, mark=*, mark size=2.5pt,
    mark options={fill=mypink}, only marks,
] coordinates {(12.5, 8.26)};
\node[mypink, anchor=north west, font=\small] at (axis cs:12.5, 8.3) {FocalCodec};
 
\addplot[
    color=myolive, mark=*, mark size=2.5pt,
    mark options={fill=myolive}, only marks,
] coordinates {(12.5, 3.95)};
\node[myolive, anchor=south west, font=\small] at (axis cs:12.5, 4.0) {Kanade};
 
\end{axis}
\end{tikzpicture}
\label{fig:wer}

\end{minipage}

\caption{Left: SIM vs. Token rate. Right: WER vs. Token rate. \modelname~substantially improves over existing codecs with similar token rate.}
\end{center}
\end{figure*}

\section{Introduction}

The rise of Large Language Model (LLM)-based audio generation, latent diffusion-based Text-to-Speech (TTS), and multimodal speech integration has driven an urgent need for efficient discrete acoustic representations. A key component of these pipelines is the neural audio codec, which compresses speech into token sequences that downstream models can process. These tokens must simultaneously preserve acoustic and semantic information for effective downstream task performance while maintaining high reconstruction quality.

A key gap between the audio and text modalities is tokenization efficiency. Unlike raw text, audio waveforms of speech contain non-linguistic information such as speaker identity, background noise, and other acoustic information. As a reuslt, audio codecs require significantly higher token rates than text tokenizers - often hundreds of tokens per second of audio. This presents significant challenges towards LLM-based audio approaches, as even modestly long audio requires thousands of tokens to model, making it challenging to capture long-term dependencies and increasing computational costs. Encoding this information using as few tokens as possible remains a primary focus to improve the efficiency of downstream audio tasks.

Despite recent progress, existing audio codecs face challenges when balancing reconstruction quality with token efficiency. Semantic codecs utilize features from self-supervised models like WavLM \citep{DBLP:journals/jstsp/ChenWCWLCLKYXWZ22}, HuBERT \citep{DBLP:journals/taslp/HsuBTLSM21}, and wav2vec2 \citep{DBLP:conf/nips/BaevskiZMA20}, efficiently capturing linguistic information. However, these approaches fail to encode non-linguistic information like timbre and prosody, causing low-quality audio reconstruction. Acoustic codecs employ VQ-VAE \citep{DBLP:conf/nips/OordVK17} architectures for high-fidelity reconstruction, but they often use multiple codebooks, causing high token rates. Meanwhile, disentangled codecs separate audio into time-invariant global representations and time-dependent tokens, achieving low token rates but struggling to faithfully recreate the original audio, especially on out-of-domain distributions.

In this work, we reframe speech tokenization as a \textit{selective information bottleneck} problem. We argue that rather than treating audio compression as a task of maximally preserving all signal information, the ideal audio codec should only encode salient information and discard perceptually insignificant features such as background noise and recording artifacts. To operationalize this principle, we introduce \modelname, a \textit{denoising} audio codec that learns to encode only perceptually important features. We achieve this by proposing a novel joint training objective that couples standard audio reconstruction with a speech enhancement task that removes unimportant features before encoding, in addition to conditioning objectives that promote the encoding of linguistic and timbre information. With a token rate of just 12.5 tokens per second, \modelname~achieves SOTA tokenization efficiency while preserving semantic and timbre information more effectively than existing codecs.

Our contributions are as follows:

\begin{itemize}
    \item We propose a novel codec training framework that jointly optimizes for audio reconstruction and speech enhancement, ensuring that only salient information is preserved.
    \item We introduce \modelname, a denoising audio codec trained using our framework. According to benchmarks, \modelname~achieves superior reconstruction for a given information budget compared to existing audio codecs, demonstrating superior encoding efficiency.
    \item We demonstrate how \modelname~enables improved robustness in downstream tasks such as speech recognition and TTS.
\end{itemize}

\section{Related Work}

Existing approaches to speech tokenization broadly fall into two families. Semantic tokenizers use SSL models like HuBERT \citep{DBLP:journals/taslp/HsuBTLSM21} and wav2vec2 \citep{DBLP:conf/nips/BaevskiZMA20} map audio to linguistically meaningful representations via self-supervised learning (SSL) and k-means clustering, capturing phonetic structure well but discarding prosodic and acoustic detail. Acoustic codecs \citep{DBLP:journals/tmlr/DefossezCSA23,DBLP:journals/corr/abs-2301-02111,DBLP:conf/iclr/ZhangZLZQ24,DBLP:conf/interspeech/LiLLHWWZ025} instead train encoder-decoder networks with reconstruction losses and vector quantization (VQ) \citep{gray1984vector} to discretize audio into compact code vectors. While acoustic codecs achieve high-fidelity reconstruction, they typically rely on multi-codebook designs such as residual vector quantization (RVQ) \citep{DBLP:journals/sensors/ChenGW10} — for instance, DAC \citep{DBLP:conf/nips/KumarSLKK23} employs 9 RVQ layers — which adds substantial complexity to downstream modeling and results in token rates far exceeding text-focused tokenizers.

Single-codebook designs have emerged as a promising alternative, streamlining both training and inference while enabling more seamless integration with text-based language models \citep{DBLP:conf/iclr/ParkerSPCZEL25}. WavTokenizer \citep{DBLP:conf/iclr/Ji00C0Z0C0LZY0J25} achieves 40 tokens per second using k-means clustering and random awakening strategies, and Single-Codec \citep{DBLP:conf/interspeech/LiXGZLXCYL24} leverages a specialized BiLSTM \citep{hochreiter1997long} architecture to preserve reconstruction quality with a single quantizer. More recently, FocalCodec \citep{DBLP:journals/corr/abs-2502-04465} introduces a focal modulation-based compressor-quantizer-decompressor architecture operating at ultra-low token rates (12.5 - 50 t/s) with a single binary codebook. However, our experiments show that these single-codebook approaches continue to struggle to balance compression with both reconstruction quality and effective semantic representation, particularly at low token rates.

A parallel line of work has pursued disentangled codecs \citep{DBLP:conf/icml/JuWS0XYLLST000024,DBLP:journals/corr/abs-2404-02702,DBLP:conf/icassp/RenWYX0ZZ24}, which aim to separate linguistic content from non-linguistic information such as speaker identity and background noise. Kanade \citep{DBLP:journals/corr/abs-2602-00594} uses SSL features as encoder inputs alongside a narrow information bottleneck to achieve clean, unsupervised disentanglement. By operating on SSL features, Kanade inherits a structured latent space in which content and speaker information are easily separable, yielding strong lexical availability and prosody preservation with a single token stream. Nevertheless, using SSL features as inputs limits generalization to out-of-distribution and unseen acoustic domains, and the use of recurrent or attention-based sequence modeling constrains the codec to the input lengths seen during training.

\begin{figure}[h]
    \centering
    \includegraphics[width=0.92\linewidth]{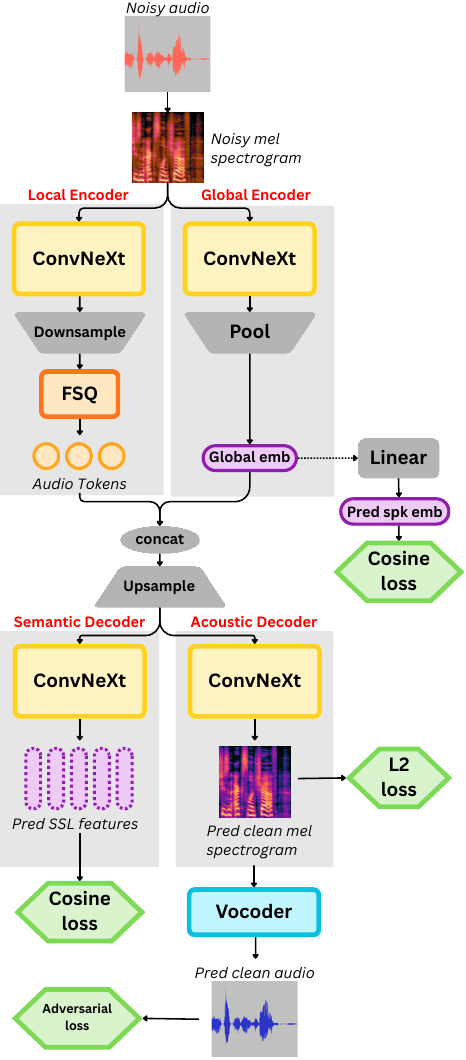}
    \caption{Architecture of \modelname, along with loss sources used during training.}
    \label{fig:architecture}
\end{figure}

\section{Methodology}

\subsection{Architecture}

Our architecture is illustrated in Figure \ref{fig:architecture}. \modelname~is comprised of (1) a disentangled VQVAE-style autoencoder operating on log-mel spectrograms and (2) a Vocos vocoder \citep{DBLP:conf/iclr/Siuzdak24} to convert mel spectrograms into raw audio. When encoding audio, \modelname~converts the audio waveform to a mel spectrogram, reconstructs the mel spectrogram after encoding it in a discrete token form, then converts the mel spectrogram back into a waveform. Notably, this avoids relying on SSL features as input, which can lead to loss of non-linguistic information and a lack of robustness. We argue that mel features are also superior to using raw audio waveforms as input, as used by HH-Codec \citep{DBLP:journals/corr/abs-2507-18897} and VibeVoice \citep{DBLP:journals/corr/abs-2508-19205}, because mel spectrograms align more closely with human perception of audio. Converting the input audio to a mel spectrogram reduces the temporal resolution from 24 kHz to 62.5 Hz.

\textbf{Local Encoder.} We adopt a fully convolutional architecture, using 1D ConvNeXt \citep{DBLP:conf/cvpr/0003MWFDX22} blocks as proposed in Vocos \citep{DBLP:conf/iclr/Siuzdak24}. After passing through the local encoder, encoder outputs are further downsampled by a factor of 5 to reach the desired token rate of 12.5 t/s. Then, we use finite scalar quantization (FSQ) \citep{DBLP:conf/iclr/MentzerMAT24} to quantize encoder outputs. FSQ is single-codebook, parameter-free, and avoids codebook divergence caused by traditional VQ. We use codebook levels of [8, 8, 8, 8 ,8] for a total of 32,768 possible codes.

\textbf{Global Encoder.} Our global encoder architecture is inspired by Kanade \citep{DBLP:journals/corr/abs-2602-00594}, using modified ConvNeXt blocks from NeXt-TDNN \citep{DBLP:conf/icassp/HeoSLCP24}. After passing through the global encoder, we then apply attentive statistics pooling \citep{DBLP:conf/interspeech/OkabeKS18} to obtain a single global embedding for the entire utterance. We use a global embedding size of 256, which is concatenated to the audio tokens during the decoding phase.

\textbf{Semantic-Acoustic Dual Decoder.} Audio codecs are frequently trained to encode both acoustic and semantic features to best preserve linguistic and non-linguistic information. We achieve this by employing a dual-decoder architecture. After receiving audio tokens augmented with the global embedding, we upsample to recover the original temporal resolution of 62.5 Hz. Then, the path of computation splits into two branches. The \textit{acoustic decoder} reconstructs the original mel spectrogram, while the \textit{semantic decoder} reconstructs SSL features produced by WavLM-large \citep{DBLP:journals/jstsp/ChenWCWLCLKYXWZ22}, a self-supervised audio model with rich semantic features. This ensures that our token representation contains both semantic and acoustic information, leading to high-quality reconstruction. Similarly to the local encoder, we use 1D ConvNeXt blocks for both decoders.

\textbf{Vocoder.} Finally, the reconstructed mel spectrogram is then passed through a Vocos-based vocoder \citep{DBLP:conf/iclr/Siuzdak24} to convert it to an audio waveform. The vocoder is an important aspect of the codec, as it directly influences reconstruction quality. Therefore, we train our own Vocos model with increased depth and width for better reconstruction.

\subsection{Training Framework}

\subsubsection{Motivation}
As token rate decreases, it becomes increasingly challenging to accurately reconstruct audio. This is because audio is composed of a mixture of \textit{perceptually important} information, such as linguistic content and speaker timbre, as well as \textit{perceptually unimportant} information, such as background noise, reverberation, or microphone quality. Existing audio codecs attempt to preserve all information in the original audio, typically by reconstructing the mel spectrogram or raw waveform. However, we argue that some aspects of audio are more important than others. The ideal audio codec should ignore all perceptually unimportant information and allocate its limited bitrate on encoding only perceptually important components of audio, since this will maximize the perceived quality of the reconstructed audio. With this principle, we develop several training strategies to promote the encoding of only relevant speech information.

\subsubsection{Speech enhancement}
In addition to training with a standard audio reconstruction task, we jointly use a speech enhancement training objective. During training, we corrupt speech utterances with a variety of audio degradations and supervise the model to reconstruct the original clean signal. This causes our model to ignore and remove perceptually unimportant information added by these degradations, allowing it to prioritize encoding important linguistic and auditory information. We use a degradation pipeline based on \citet{DBLP:journals/corr/abs-2509-17052}, with each degradation applied with a probability of 50\%:
\begin{enumerate}
    \item Reverberation: We generate random RT60 and rectangular dimensions from $\mathcal{U}(0.1,2.0)$ seconds and $\mathcal{U}(2, 20)$ m, respectively. Then, we use pyroomacoustics \citep{DBLP:conf/icassp/ScheiblerBD18} to simulate room impulse responses (RIRs). We filter any RIRs of size $>50000$ samples to avoid excessive reverberation.
    \item Background noise: We randomly sample a noise recording from a dataset pool of AudioSet \citep{DBLP:conf/icassp/GemmekeEFJLMPR17}, FSD50K \citep{DBLP:journals/taslp/FonsecaFPFS22}, and WHAM! \citep{DBLP:conf/interspeech/WichernAFZMCMR19}. We add this to the clean utterance with SNR sampled from $\mathcal{U}(15, 30)$ dB.
    \item Low-pass filter: We use a biquad lowpass filter with a cutoff frequency ranging from 2-8 kHz.
    \item Resampling: The utterance is resampled to one of $\{16, 22.05\}$ kHz before being resampled back to 24 kHz.
    \item Codec: We applied MP3 compression with a random bitrate between 32-245 kbps.
\end{enumerate}

\subsubsection{Global Conditioning}

Conditioning model outputs on SSL features is a frequently-employed technique to better preserve linguistic information. However, we believe that speaker timbre is also important to preserve. Thus, we design a novel conditioner to facilitate the encoding of speaker information. Recall that disentangled codecs attempt to encode linguistic information in the local encoder and non-linguistic information like timbre in the global encoder. As a result, we condition the global encoder using a pretrained speaker verification (SV) model. This enables stronger gradient signals to reach the global encoder, enabling better preservation of speaker timbre information.

Specifically, we use TitaNet-large \citep{DBLP:conf/icassp/KoluguriPG22}, a self-supervised speaker recognition model trained on over 100,000 speakers. We apply a learned linear projection from our global embedding to TitaNet's embedding space and train the encoder to to maximize its cosine similarity with the pretrained embeddings. To our knowledge, this is the first time speaker recognition models have been used to condition audio codecs for improved speaker reconstruction.

\subsubsection{Training objectives}

We use the following notation:

\begin{itemize}
    \item $m$: true log-mel spectrogram of utterance
    \item $s$: true SSL features from WavLM-large
    \item $g$: true speaker embedding from TitaNet-large
    \item $\hat{g}$: pred embedding from global encoder
    \item $\hat{m}$: pred log-mel spectrogram from acoustic decoder
    \item $\hat{s}$: pred SSL features from semantic decoder
\end{itemize}

The autoencoder uses three loss sources:
\begin{enumerate}
    \item L2 loss between the original mel spectrogram and the output produced by the acoustic decoder, denoted $\mathcal{L}_\text{mel}$:
    $$\mathcal{L}_\text{mel}=||m-\hat{m}||_2$$
    \item Cosine loss between the semantic representation produced by WavLM-large and the output produced by the semantic decoder, denoted $\mathcal{L}_\text{ssl}$. Formally, assuming $\cos(x,y)$ is the cosine similarity between $x$ and $y$,
    $$\mathcal{L}_\text{ssl}=1-\cos(s,\hat{s})$$
    \item Cosine loss between the speaker embedding produced by TitaNet-large and the output from the global encoder, denoted $\mathcal{L}_\text{emb}$. Assuming $p$ is a learned linear projection,
    $$\mathcal{L}_\text{emb}=1-\cos(g,p(\hat{g}))$$
\end{enumerate}

Vocoder training uses the adversarial training recipe from Vocos \citep{DBLP:conf/iclr/Siuzdak24}. We use a multi-period discriminator \citep{DBLP:conf/nips/KongKB20} and a multi-band STFT discriminator at multiple timescales \citep{DBLP:conf/interspeech/JangLYKK21}. The vocoder loss is composed of multiscale mel spectrogram loss \citep{DBLP:conf/iclr/LeePGCY23}, adversarial loss, and feature matching loss.

We find that training the autoencoder and vocoder together from scratch leads to bad performance because the high reconstruction error early in training leads to adversarial training collapse, corroborating results found in \citet{DBLP:journals/corr/abs-2507-18897}. Instead, we adopt a two stage framework. In the first stage, we train the autoencoder and vocoder independently. Critically, the bottleneck-contrained autoencoder is trained without adversarial training, avoiding training divergence. After both models have converged, we then fine-tune both models in an end-to-end fashion. We find this vastly improves training stability and reconstruction quality.

\section{Experiments}

\subsection{Training Setup}

We train \modelname~on the full training set of LibriTTS-R \citep{DBLP:conf/interspeech/KoizumiZKDYMB0H23}, consisting of 585 hours of read speech from 2,456 speakers, as well as a 1,800-hour subset from the Emilia-YODAS dataset \citep{DBLP:journals/corr/abs-2501-15907}, bringing the total dataset size to approximately 2,400 hours. To evaluate the scaling performance of \modelname, we train model variants with a downsampling factor of 1 and 2 (as opposed to 5), achieving token rates of 62.5 t/s and 31.25 t/s, respectively. We refer to these models as \modelname@62.5, \modelname@31.25, and \modelname@12.5. Additional training details can be found in Appendix \ref{app:hyperparameters}.

\subsection{Baselines}

We compare our method with a variety of recent SOTA audio codecs, including Mimi \citep{DBLP:journals/corr/abs-2410-00037}, BiCodec \citep{DBLP:journals/corr/abs-2503-01710}, XCodec2 \citep{DBLP:journals/corr/abs-2502-04128}, WavTokenizer \citep{DBLP:conf/iclr/Ji00C0Z0C0LZY0J25}, FocalCodec \citep{DBLP:journals/corr/abs-2502-04465}, Qwen3-TTS-Tokenizer (Qwen3) \citep{DBLP:journals/corr/abs-2601-15621}, and Kanade \citep{DBLP:journals/corr/abs-2602-00594}. 

\subsection{Evaluation}

We evaluate reconstructed audio on:
\begin{itemize}
    \item \textbf{Speaker identity:} Speaker embedding cosine similarity (SIM) using ReDimNet-M speaker verification model \citep{DBLP:conf/interspeech/YakovlevMBMOT24}. We use ReDimNet-M instead of the more common WavLM-SV because it is a stronger SV model and provides greater discriminability between baselines.
    \item \textbf{Reconstruction faithfulness:} L1 distance between log-mel spectrograms (Mel L1).
    \item \textbf{Speech intelligibility:} word/character error rate (WER/CER) using \texttt{parakeet-tdt-0.6B-v2} speech recognition model \citep{DBLP:journals/corr/abs-2509-14128}. These metrics are computed relative to the Parakeet transcription of the original audio so that the ground truth has a WER/CER of 0.
    \item \textbf{Quality:} UTMOS \citep{DBLP:conf/interspeech/SaekiXNKTS22}, SI-SDR from the Torchaudio-Squim evaluation suite \citep{DBLP:conf/icassp/KumarTNMZHX23}.
\end{itemize}

\begin{table*}[h]
\centering
\resizebox{\textwidth}{!}{%
\begin{tabular}{lc|cccccc|cccccc}
\toprule
 & & \multicolumn{6}{c|}{LibriTTS test-clean} & \multicolumn{6}{c}{LibriTTS test-other} \\
 & Token Rate   & WER & CER & SIM & UTMOS & SI-SDR & Mel L1
                & WER & CER & SIM & UTMOS & SI-SDR & Mel L1 \\
 & (t/s) & ($\downarrow$) & ($\downarrow$) & ($\uparrow$) & ($\uparrow$) & ($\uparrow$) & ($\downarrow$)
         & ($\downarrow$) & ($\downarrow$) & ($\uparrow$) & ($\uparrow$) & ($\uparrow$) & ($\downarrow$) \\
\midrule
GT & --             & 0.0 & 0.0 & 1.00 & 4.20 & 26.49 & 0.00
                    & 0.0 & 0.0 & 1.00 & 4.01 & 24.44 & 0.00 \\
\midrule
\multicolumn{14}{c}{\textit{Token Rate $\mathit{>100}$ t/s}} \\
\midrule
Qwen3               & 200   & 1.3       & 0.7       & 0.90      & 4.27      & 27.99     & 0.29
                            & 2.6       & 1.4       & 0.90      & 4.09      & 27.85     & 0.29 \\
\midrule
\multicolumn{14}{c}{\textit{Token Rate $\mathit{\ge50}$ t/s}} \\
\midrule
Mimi       & 100            & 3.5       & 1.9       & 0.72      & 3.86      & 17.48     & 0.56     
                            & 7.5       & 4.4       & 0.70      & 3.63      & 17.21     & 0.56 \\
Qwen3               & 100   & 2.6       & 1.4       & 0.72      & 3.92      & 19.45     & 0.38
                            & 4.7       & 2.5       & 0.72      & 3.73      & 20.20     & 0.38 \\
BiCodec    & 50             & 2.6       & 1.3       & 0.78      & \U{4.25}  & \B{26.49} & 0.55     
                            & 6.2       & 3.4       & 0.77      & \U{4.07}  & \B{26.14} & 0.62 \\
XCodec2    & 50             & 2.6       & 1.5       & 0.77      & 4.20      & \U{23.51} & 0.39 
                            & 5.4       & 3.1       & 0.75      & 4.01      & \U{23.47} & 0.40 \\
Qwen3               & 50  &   7.9       & 4.5       & 0.31      & 2.65      & 9.01      & 0.63
                            & 15.8      & 9.4       & 0.30      & 2.51      & 8.91      & 0.61 \\
\midrule    
\modelname@62.5&62.5        & \B{1.3}   & \B{0.6}   & \B{0.90}  & 4.23      & 22.94     & \B{0.29}     
                            & \B{2.7}   & \B{1.6}   & \B{0.89}  & 4.06      & 22.83     & \B{0.29} \\
\modelname@31.25  & 31.25   & \U{1.7}   & \U{0.9}   & \U{0.88}  & \B{4.27}  & 22.02     & \U{0.33}
                            & \U{3.6}   & \U{1.9}   & \U{0.87}  & \B{4.10}  & 22.12     & \U{0.33} \\
\midrule
\multicolumn{14}{c}{\textit{Token Rate $\mathit{<50}$ t/s}} \\
\midrule
WavTokenizer & 40           & 9.0       & 5.0       & 0.62      & 3.81      & 19.98     & \U{0.46}
                            & 19.1      & 11.5      & 0.61      & 3.63      & 20.94     & \U{0.47}  \\
FocalCodec & 12.5           & 8.3       & 4.5       & 0.45      & 4.21      & \B{25.19} & 0.78
                            & 16.8      & 9.7       & 0.42      & 4.05      & \B{25.25} & 0.79  \\
Kanade     & 12.5           & \U{4.0}   & \U{2.1}   & \U{0.65}  & \U{4.22}  & \U{23.60} & 0.57     
                            & \U{8.2}   & \U{4.5}   & \U{0.64}  & \U{4.10}  & \U{23.77} & 0.59 \\
\midrule
\modelname@12.5    & 12.5   & \B{2.7}   & \B{1.4}   & \B{0.86}  & \B{4.32}  & 21.96     & \B{0.44}
                            & \B{5.7}   & \B{3.2}   & \B{0.84}  & \B{4.17}  & 21.73     & \B{0.44} \\
\bottomrule
\end{tabular}
}
\caption{Evaluation results on LibriTTS test-clean and test-other.}
\label{tab:main}
\end{table*}

We also report the token rate of each audio codec. Smaller token rates compress audio more aggressively, so codecs with higher token rates should have better metrics. As a result, we classify our baselines by token rate, and compare our method against baselines of similar token rate. Specifically, we compare \modelname@12.5 with baselines below 50 t/s, and compare \modelname@62.5 and \modelname@31.25 with baselines above 50 t/s. We take care to ensure that our models generally have a lower token rate than the baselines they are comparing to so that our improvements cannot be confounded with using a higher token rate.

\subsection{Speech Reconstruction}

We evaluate all baselines on the test-clean and test-other splits of LibriTTS \citep{DBLP:conf/interspeech/ZenDCZWJCW19}, containing 8.6 and 6.7 hours, respectively. Our results can be found in Table \ref{tab:main}, with bold strongest results and underlined second-strongest. \modelname@12.5~achieves either the first or second-strongest results in the vast majority of benchmarks, achieving a new SOTA result. Notably, \modelname@12.5~improves substantially over existing baselines in SIM (0.86 vs 0.64) and WER (2.7 vs 4.0) scores, even achieving competitive results with other codecs with token rates $>$50 t/s. This indicates that \modelname@12.5~achieves much better speaker reconstruction and speech intelligibility compared to other audio codecs.

Extending our methodology to higher token rates, we see that \modelname's performance consistently improves as token rate increases. \modelname@31.25 surpasses both BiCodec and XCodec2 in 5 of 6 benchmarks, despite these codecs using a 60\% higher token rate. In addition, \modelname@62.5 achieves competitive performance with Qwen3-TTS-Tokenizer while having a 3x lower token rate.

\begin{table}[h]
\centering
\resizebox{0.95\linewidth}{!}{%
\begin{tabular}{lc|cccccc}
\toprule
 & Token Rate & WER & CER & SIM & UTMOS & SI-SDR & Mel L1 \\
 & (t/s) & ($\downarrow$) & ($\downarrow$) & ($\uparrow$) & ($\uparrow$) & ($\uparrow$) & ($\downarrow$) \\
\midrule
\multicolumn{8}{c}{\textit{\textbf{Expresso}}} \\
\midrule
GT & --           & 0.0   & 0.0   & 1.00  & 3.43  & 24.15 & 0.00  \\
\midrule
WavTokenizer & 40           & 19.7      & 11.4      & 0.49      & 2.95      & 16.56     & \U{0.53}  \\
FocalCodec & 12.5           & 21.0      & 11.3      & 0.34      & \B{3.75}  & 22.53     & 0.89      \\
Kanade     & 12.5           & \U{9.3}   & \U{4.9}   & \U{0.55}  & 3.38      & \U{22.81} & 0.61      \\
\midrule
\modelname@12.5 & 12.5      & \B{3.9}   & \B{1.9}   & \B{0.82}  & \U{3.55}  & \B{22.99} & \B{0.45}  \\
\midrule
\multicolumn{8}{c}{\textit{\textbf{AISHELL-3}}} \\
\midrule
GT & --           & --    & 0.0   & 1.00  & 2.68  & 21.78 & 0.00  \\
\midrule
WavTokenizer & 40           & --        & \U{6.1}   & \U{0.50}  & 2.45      & 18.90     & \U{0.52}  \\
FocalCodec & 12.5           & --        & 15.0      & 0.24      & \B{3.68}  & \B{25.43} & 0.79  \\
Kanade     & 12.5           & --        & 7.5       & 0.47      & \U{3.18}  & \U{24.45} & 0.58  \\
\midrule
\modelname@12.5 & 12.5      & --        & \B{1.5}   & \B{0.84}  & 3.04      & 22.41     & \B{0.39}  \\
\midrule
\multicolumn{8}{c}{\textit{\textbf{CML-TTS}}} \\
\midrule
GT & --           & 0.0   & 0.0   & 1.00  & 3.03  & 22.55 & 0.00  \\
\midrule
WavTokenizer & 40           & 29.6      & 14.8      & \U{0.58}  & 2.71      & 20.51     & \B{0.47}  \\
FocalCodec & 12.5           & 54.4      & 29.9      & 0.35      & \B{3.58}  & \B{24.94} & 0.75  \\
Kanade     & 12.5           & \U{42.1}  & \U{21.4}  & 0.57      & 3.42      & \U{23.98} & \U{0.59}  \\
\midrule
\modelname@12.5 & 12.5      & \B{12.1}  & \B{5.1}   & \B{0.78}  & \U{3.47}  & 22.11     & 0.65  \\
\bottomrule
\end{tabular}%
}
\caption{Evaluation results on OOD speech. Full results in Table \ref{tab:robustness}.}
\label{tab:robustness-sample}
\end{table}

To demonstrate the robustness of our model, we also evaluate using a variety of out-of-domain datasets, including Expresso \citep{DBLP:conf/interspeech/NguyenHDSGFRCSH23}, AISHELL-3 \citep{DBLP:journals/corr/abs-2010-11567}, and CML-TTS \citep{DBLP:conf/slt/ConneauMKZADRRB22}, covering expressive, multilingual, and in-the-wild speech. Table \ref{tab:robustness-sample} contains OOD results, with full results in Appendix \ref{app:robustness} due to page limit constraints. Despite not explicitly being trained for any of these domains, we find that \modelname~achieves strong results compared to other audio codecs, maintaining or improving its lead against baselines. In particular, we find that audio codecs with low token rates exhibit significantly degraded speech intelligibility on OOD datasets. However, \modelname@12.5 experiences much less degradation, achieving competitive WER/CER in Expresso and AISHELL-3.

\subsection{Disentanglement Effectiveness}

Disentangled codecs aim to separate audio into speaker characteristics (located in the global representation) and the linguistic content (located in the local representation). To measure the disentanglement performance, we train models on (1) speaker verification performance using only the global representation and (2) speech recognition performance using only the local representation. For the speaker verification task, we directly use the global representation as the speaker embedding. We conduct speaker verification experiments using VoxCeleb-1 \citep{DBLP:conf/interspeech/NagraniCZ17}. We evaluate using equal error rate (EER) and accuracy (ACC) based on the closest speaker centroid. For speech recognition, we train decoder-only transformers on local audio tokens from LibriTTS. We evaluate using the LibriTTS test-clean split with WER and CER. We compare \modelname~with our disentangled baselines: BiCodec and Kanade. We do not test against older disentangled codecs like FACodec \citep{DBLP:conf/icml/JuWS0XYLLST000024} and TiCodec \citep{DBLP:conf/icassp/RenWYX0ZZ24} because they have been shown to have noncompetitive performance compared to our baselines.

Our results can be found in Table \ref{tab:disentangled}. Our models achieve SOTA results in both tasks, indicating that \modelname~separates linguistic and timbre information the best compared to our baselines. We attribute this to our global and local encoders with TitaNet and WavLM conditioning, respectively.

\begin{table}[h]
\centering
\resizebox{\linewidth}{!}{%
\begin{tabular}{lc|cc|cc}
\toprule
 & & \multicolumn{2}{c|}{SV} & \multicolumn{2}{c}{ASR} \\
 & Token Rate & ACC & EER & WER & CER\\
 & (t/s)      & ($\uparrow$) & ($\downarrow$) & ($\downarrow$) & ($\downarrow$) \\
\midrule
BiCodec         & 50    & 99.15 & 1.52 & 6.2 & 3.4 \\
Kanade          & 12.5  & 96.82 & 3.38 & 8.1 & 4.0 \\
\midrule
\modelname@62.5 & 62.5  & \B{99.99} & \U{0.23} & \B{4.7} & \B{2.6} \\
\modelname@31.25& 31.25 & \U{99.98} & \B{0.22} & \U{4.8} & \U{2.7} \\
\modelname@12.5 & 12.5  & 99.92     & 0.58     & 5.6     & 3.0 \\
\bottomrule
\end{tabular}%
}
\caption{SV and ASR results for disentangled codecs.}
\label{tab:disentangled}
\end{table}

\subsection{Downstream Tasks}

We evaluate our model on two downstream tasks:

\begin{enumerate}
    \item \textbf{Voice conversion (VC):} Given source and reference audio clips, convert the source clip to the reference voice while preserving the source linguistic content. As shown in \citet{DBLP:journals/corr/abs-2602-00594}, non-disentangled codecs significantly underperform at VC tasks, so we compare our method against BiCodec and Kanade. To obtain voice-converted audio, we extract global and local representations from the source and reference audio. Then, we pass the global embedding from the reference audio and the local tokens from the source audio through the decoder to obtain the converted audio. We evaluate using VCTK \citep{Yamagishi2019-xf}, a 40-hour dataset of parallel utterances from 109 speakers, using WER/CER (linguistic similarity), SIM (speaker similarity), and UTMOS (quality).
    \item \textbf{Text-to-speech (TTS):} Given a reference audio and a text, generate audio of the reference voice speaking the given text. We train decoder-only transformers on local audio tokens from LibriTTS. We evaluate using Seed-TTS-eval \citep{DBLP:journals/corr/abs-2406-02430}, and also report the training time and inference speed.
\end{enumerate}

\begin{table}[h]
\centering
\resizebox{\linewidth}{!}{%
\begin{tabular}{lc|cccc}
\toprule
 & Token Rate & SIM & UTMOS & WER & CER\\
 & (t/s)      & ($\uparrow$) & ($\uparrow$) & ($\downarrow$) & ($\downarrow$) \\
\midrule
GT    & --    & 1.00  & 4.10 & 0.0 & 0.0 \\
\midrule
BiCodec         & 50    & 0.47      & 3.87      & 1.3   & 0.7 \\
Kanade          & 12.5  & 0.52      & \B{4.21}  & 2.2   & 1.1 \\
\midrule
\modelname@62.5 & 62.5  & 0.70      & 3.93      & \B{0.8}   & \B{0.3} \\
\modelname@31.25& 31.25 & \U{0.76}  & 3.95      & \U{0.9}   & \U{0.4} \\
\modelname@12.5 & 12.5  & \B{0.81}  & \U{4.09}  & 1.4       & 0.7   \\
\bottomrule
\end{tabular}%
}
\caption{VC results for disentangled codecs on VCTK.}
\label{tab:vc}
\end{table}

VC results can be found in Table \ref{tab:vc}. \modelname~achieves the highest speaker similarity by a wide margin (0.81 for \modelname@12.5 vs 0.52 for Kanade). Although BiCodec achieved strong speaker similarity in the speech reconstruction task, we observed that changing the global encoding rarely changed the speaker's identity, indicating that BiCodec stores significant speaker information in the local audio tokens. As a result, it performed poorly on this task. \modelname~also appears to exhibit this behavior, albeit to a lesser degree, as both speaker similarity and UTMOS scores moderately degrade as token rate increases. Nevertheless, it still achieves SOTA results in 3 of 4 metrics. TTS results can be found in Table \ref{tab:tts}. Here, the efficiency advantages of using a lower token rate become clear. Both training and inference speed are inversely proportional to the token rate, allowing a TTS model using \modelname~to have 10x faster training and 17x faster inference compared to using Qwen3-TTS-Tokenizer. In addition, using a lower token rate shortens training sequences, allowing better modeling of audio. These factors combine for \modelname~to achieve SOTA results compared to baselines, both in terms of efficiency and performance. This indicates that the resulting audio tokens produced by \modelname~contains rich information for speech synthesis.

\begin{table}[h]
\centering
\resizebox{\linewidth}{!}{%
\begin{tabular}{lc|cc|cc}
\toprule
 & Token Rate & Train Time & RTF (RTFx) & SIM & WER \\
 & (t/s)      & (hh:mm, $\downarrow$) & ($\downarrow,\uparrow$) & ($\uparrow$) & ($\downarrow$) \\
\midrule
Qwen3           & 200   & 10:02     & 2.930 (0.3x)  & 0.36      & 9.1 \\
Mimi            & 100   & 05:11     & 1.452 (0.7x)  & 0.34      & 10.8 \\
XCodec2         & 50    & 02:44     & 0.688 (1.5x)  & 0.40      & \U{4.9} \\
BiCodec         & 50    & 02:49     & 0.691 (1.4x)  & \U{0.51}  & 5.2 \\
WavTokenizer    & 40    & 02:19     & 0.552 (1.8x)  & 0.37      & 16.6 \\
FocalCodec      & 12.5  & 00:58     & 0.172 (5.8x)  & 0.31      & 12.2 \\
Kanade          & 12.5  & 00:59     & 0.169 (5.9x)  & 0.45      & 5.6 \\
\modelname      & 12.5  & 01:00     & 0.170 (5.9x)  & \B{0.56}  & \B{3.9} \\
\bottomrule
\end{tabular}%
}
\caption{TTS results on Seed-TTS-eval.}
\label{tab:tts}
\end{table}

\subsection{Efficiency Analysis}
We also conduct a thorough efficiency analysis of \modelname, measuring its inference speed as well as how efficiently it utilizes its codebook. We move this analysis to Appendix \ref{app:efficiency} due to page limit constraints.

\subsection{Ablation Study}

We conduct an extensive ablation study to validate the design choices used, with results in Table \ref{tab:ablation}:

\begin{table}[h]
\centering
\resizebox{\linewidth}{!}{%
\begin{tabular}{c|cccccc}
\toprule
 & WER & CER & SIM & UTMOS & SI-SDR & Mel L1 \\
 & ($\downarrow$) & ($\downarrow$) & ($\uparrow$) & ($\uparrow$) & ($\uparrow$) & ($\downarrow$) \\
\midrule
Baseline                    & 2.7   & 1.4   & 0.86  & 4.32  & 21.96     & 0.44 \\
\midrule
One-stage training          & 4.1   & 2.2   & 0.80  & 4.01  & 18.58     & 0.57 \\
HuBERT SSL cond.            & 2.9   & 1.4   & 0.85  & 4.32  & 22.03     & 0.45 \\
No SSL cond.                & 8.1   & 4.3   & 0.84  & 4.34  & 22.56     & 0.42 \\
WavLM global cond.          & 2.9   & 1.5   & 0.77  & 4.32  & 21.55     & 0.44 \\
No global cond.             & 2.8   & 1.5   & 0.74  & 4.33  & 21.88     & 0.43 \\
No denoise                  & 3.4   & 1.7   & 0.82  & 4.24  & 20.55     & 0.41 \\
\bottomrule
\end{tabular}%
}
\caption{Ablation study.}
\label{tab:ablation}
\end{table}

\begin{enumerate}
    \item \textbf{Two-stage training:} We jointly train the autoencoder and vocoder from scratch, instead of training each separately followed by end-to-end fine-tuning. This causes significant degradations in all metrics. Early during training, we find the model struggles to fool the discriminators, leading to high early reconstruction loss that never recovers after training.
    \item \textbf{SSL conditioning:} We experiment with removing the SSL reconstruction branch from our decoder. Although there is a slight improvement in non-linguistic quality metrics like UTMOS and SI-SDR, there is a sharp drop in linguistic metrics, with a 3x increase in WER and CER. We also replace HuBERT-large with WavLM-large as the pretrained SSL model. We see that performance is very similar, indicating that \modelname~is not sensitive to the choice of SSL model.
    \item \textbf{Global conditioning:} We remove the global conditioning loss from the TitaNet-large speaker verification model. This causes a drop in speaker similarity without significant changes to other metrics. We also replace TitaNet with WavLM-SV, and find that this harms speaker reconstruction, corroborating our claim that WavLM-SV is subpar for sufficiently discriminating between speakers.
    \item \textbf{Speech enhancement:} We retrain our models without any changes to input audio. This led to a moderate degradation in 5 of 6 metrics, indicating that our speech enhancement framework is effective in prioritizing the encoding of perceptually significant information. 
\end{enumerate}

\section{Conclusion}
We introduce \modelname, a denoising disentangled audio codec that achieves state-of-the-art performance at 12.5 tokens per second by selectively encoding perceptually important audio information. Through joint speech enhancement training, SSL and speaker conditioning, and a two-stage training framework, \modelname~demonstrates superior reconstruction quality, disentanglement, and downstream task efficiency compared to existing codecs.

\section*{Limitations}
Although \modelname~achieves strong disentanglement results, the voice conversion task demonstrates that this disentanglement degrades as token rate increases. Determining how to effectively decode local and global information for codecs with high token rates remains an area of future work. In addition, it is unclear how the downstream improvements scale across different compute budgets, as the gaps between the codecs may shrink if trained for longer. Finally, we acknowledge the potential risks of advancing speech tokenization capabilities, such as facilitating deepfake technology, although we believe the benefits of improved speech modeling outweigh these risks.


\bibliography{custom}

\appendix
\section{Training Hyperparameters}
\label{app:hyperparameters}
The local encoder and acoustic decoder both use a 16-layer, 768-dim ConvNeXt backbone with a depthwise kernel size of 27 and an MLP intermediate size of 3072. The semantic decoder is identical to the acoustic decoder, but with a kernel size of 53. The global encoder is an 8-layer, 512-dim ConvNeXt backbone with a depthwise kernel size of 7 and an MLP intermediate size of 1536. The vocoder is a 16-layer, 1536-dim ConvNeXt backbone with a depthwise kernel size of 5 and an MLP intermediate size of 4608. In total, the autoencoder contains 243M parameters and the vocoder contains 228M parameters, for a total model size of 471M parameters. However, the semantic decoder can be discarded after training, reducing parameter count to 396M parameters.

During the first training stage, we independently train the autoencoder and vocoder for 1M steps each. We use audio samples of length 230,400 samples (9.6 seconds) with a batch size of 4. All utterances under this length are repeated, then trimmed to the desired length. We also apply a high-pass filter to all audio samples with a cutoff frequency of 80 Hz, as well as loudness normalization with a peak amplitude of 0.75. We optimize using AdamW $(\beta_1=0.8,\beta_2=0.9,\text{wd}=0.01)$ and a cosine learning rate schedule initialized at 1e-4. During the second stage, we decrease the learning rate to 2e-6 and jointly train the autoencoder and vocoder in an end-to-end fashion for 200K steps. All other hyperparameters remain the same. Training was conducted using 1x A100 80GB GPU for one week.

In downstream ASR and TTS tasks, we use 16-layer, 4-head, 512-dim LLaMA-style transformers with a total of 96M parameters. For ASR models, we use training sequences of the format \texttt{<audio tokens><BOS><text><EOS>}. Only text tokens are used for computing crossentropy loss. For TTS models, we use training sequences of the format \texttt{<prompt tokens><text><BOS><audio tokens><EOS>}. Only the audio tokens are used for computing crossentropy loss. For disentangled codecs, we prepend the global embedding to the beginning of the training sequence. Following \citet{DBLP:conf/nips/CopetKGRKSAD23} for multi-codebook codecs, we flatten tokens and use a combined vocabulary, as this achieves the strongest performance. In both ASR and TTS tasks, models are trained for 10K steps with a batch size of 2400 seconds. We optimize the model using AdamW $(\beta_1=0.9,\beta_2=0.95,\text{wd}=0.1)$ and a cosine learning rate schedule initialized at 1e-3. 

\section{Out-Of-Distribution Reconstruction Results}
\label{app:robustness}
See Table \ref{tab:robustness} for full OOD results.

\begin{table}[h]
\centering
\resizebox{0.95\linewidth}{!}{%
\begin{tabular}{lc|cccccc}
\toprule
 & Token Rate & WER & CER & SIM & UTMOS & SI-SDR & Mel L1 \\
 & (t/s) & ($\downarrow$) & ($\downarrow$) & ($\uparrow$) & ($\uparrow$) & ($\uparrow$) & ($\downarrow$) \\
\midrule
\multicolumn{8}{c}{\textit{\textbf{Expresso}}} \\
\midrule
GT & --           & 0.0   & 0.0   & 1.00  & 3.43  & 24.15 & 0.00  \\
\midrule
\multicolumn{8}{c}{\textit{Token Rate $\mathit{>100}$ t/s}} \\
\midrule
Qwen3               & 200   & 1.5   & 0.6   & 0.89  & 3.50  & 25.38 & 0.29  \\
\midrule
\multicolumn{8}{c}{\textit{Token Rate $\mathit{\ge50}$ t/s}} \\
\midrule
Mimi       & 100            & 5.5       & 3.0       & 0.63      & 3.07      & 15.94     & 0.58      \\
Qwen3               & 100   & 2.9       & 1.4       & 0.68      & 3.23      & 21.15     & \U{0.38}  \\
BiCodec    & 50             & 4.5       & 2.4       & \U{0.74}  & 3.42      & \B{23.94} & 0.53      \\
XCodec2    & 50             & 4.3       & 2.1       & 0.73      & 3.34      & 22.88     & 0.41      \\
Qwen3               & 50    & 13.0      & 7.3       & 0.25      & 2.22      & 10.22     & 0.65      \\
\midrule
\modelname@62.5&62.5        & \B{1.9}   & \B{0.9}   & \B{0.87}  & \U{3.43}  & \U{23.07} & \B{0.31}  \\
\modelname@31.25  & 31.25   & \U{2.4}   & \U{1.2}   & 0.86      & \B{3.47}  & 22.87     & 0.35      \\
\midrule
\multicolumn{8}{c}{\textit{Token Rate $\mathit{<50}$ t/s}} \\
\midrule
WavTokenizer & 40           & 19.7      & 11.4      & 0.49      & 2.95      & 16.56     & \U{0.53}  \\
FocalCodec & 12.5           & 21.0      & 11.3      & 0.34      & \B{3.75}  & 22.53     & 0.89      \\
Kanade     & 12.5           & \U{9.3}   & \U{4.9}   & \U{0.55}  & 3.38      & \U{22.81} & 0.61      \\
\midrule
\modelname@12.5 & 12.5      & \B{3.9}   & \B{1.9}   & \B{0.82}  & \U{3.55}  & \B{22.99} & \B{0.45}  \\
\midrule
\multicolumn{8}{c}{\textit{\textbf{AISHELL-3}}} \\
\midrule
GT & --           & --    & 0.0   & 1.00  & 2.68  & 21.78 & 0.00  \\
\midrule
\multicolumn{8}{c}{\textit{Token Rate $\mathit{>100}$ t/s}} \\
\midrule
Qwen3               & 200   & --    & 0.5   & 0.89  & 2.96  & 24.05 & 0.30  \\
\midrule
\multicolumn{8}{c}{\textit{Token Rate $\mathit{\ge50}$ t/s}} \\
\midrule
Mimi       & 100            & --        & 1.9       & 0.57      & 2.61      & 18.38     & 0.55  \\
Qwen3               & 100   & --        & 1.3       & 0.68      & 2.77      & 22.09     & 0.37  \\
BiCodec    & 50             & --        & 1.0       & 0.77      & \U{2.92}  & 22.72     & 0.55  \\
XCodec2    & 50             & --        & 1.1       & 0.76      & 2.83      & 22.79     & 0.40  \\
Qwen3               & 50    & --        & 17.8      & 0.24      & 2.02      & 9.71      & 0.63  \\
\midrule
\modelname@62.5&62.5        & --        & \B{0.5}   & \B{0.88}  & 2.89      & \B{23.06} & \B{0.29}  \\
\modelname@31.25  & 31.25   & --        & \U{0.7}   & \U{0.87}  & \B{2.97}  & \U{22.88} & \U{0.32}  \\
\midrule
\multicolumn{8}{c}{\textit{Token Rate $\mathit{<50}$ t/s}} \\
\midrule
WavTokenizer & 40           & --        & \U{6.1}   & \U{0.50}  & 2.45      & 18.90     & \U{0.52}  \\
FocalCodec & 12.5           & --        & 15.0      & 0.24      & \B{3.68}  & \B{25.43} & 0.79  \\
Kanade     & 12.5           & --        & 7.5       & 0.47      & \U{3.18}  & \U{24.45} & 0.58  \\
\midrule
\modelname@12.5 & 12.5      & --        & \B{1.5}   & \B{0.84}  & 3.04      & 22.41     & \B{0.39}  \\
\midrule
\multicolumn{8}{c}{\textit{\textbf{CML-TTS}}} \\
\midrule
GT & --           & 0.0   & 0.0   & 1.00  & 3.03  & 22.55 & 0.00  \\
\midrule
\multicolumn{8}{c}{\textit{Token Rate $\mathit{>100}$ t/s}} \\
\midrule
Qwen3               & 200   & 3.2   & 1.4   & 0.92  & 3.15  & 24.54 & 0.30  \\
\midrule
\multicolumn{8}{c}{\textit{Token Rate $\mathit{\ge50}$ t/s}} \\
\midrule
Mimi       & 100            & 11.0      & 5.1       & 0.69      & 2.67      & 16.46     & 0.55  \\
Qwen3               & 100   & 6.3       & \U{2.5}   & 0.72      & 2.74      & 20.57     & \B{0.37}  \\
BiCodec    & 50             & 8.9       & 3.8       & 0.77      & 3.17      & 22.79     & 0.48  \\
XCodec2    & 50             & \U{6.2}   & 2.6       & 0.81      & 3.18      & \B{23.16} & \U{0.38}  \\
Qwen3               & 50    & 23.4      & 11.5      & 0.30      & 1.85      & 9.32      & 0.56  \\
\midrule
\modelname@62.5&62.5        & \B{5.4}   & \B{2.4}   & \B{0.83}  & \B{3.41}  & \U{22.86} & 0.50  \\
\modelname@31.25  & 31.25   & 6.4       & 2.7       & \U{0.82}  & \U{3.33}  & 21.72     & 0.55  \\
\midrule
\multicolumn{8}{c}{\textit{Token Rate $\mathit{<50}$ t/s}} \\
\midrule
WavTokenizer & 40           & 29.6      & 14.8      & \U{0.58}  & 2.71      & 20.51     & \B{0.47}  \\
FocalCodec & 12.5           & 54.4      & 29.9      & 0.35      & \B{3.58}  & \B{24.94} & 0.75  \\
Kanade     & 12.5           & \U{42.1}  & \U{21.4}  & 0.57      & 3.42      & \U{23.98} & \U{0.59}  \\
\midrule
\modelname@12.5 & 12.5      & \B{12.1}  & \B{5.1}   & \B{0.78}  & \U{3.47}  & 22.11     & 0.65  \\
\bottomrule
\end{tabular}%
}
\caption{Full evaluation results on OOD speech.}
\label{tab:robustness}
\end{table}

\section{Efficiency Analysis}
\label{app:efficiency}

\subsection{Codebook efficiency}
We compare the codebook utilization of \modelname~using normalized entropy:
$$\text{Normalized Entropy}=-\frac1{\log N}\sum_{x=1}^N p(x)\log(p(x)).$$
Results are reported in Table \ref{tab:codebook}. Our model has the second-highest normalized entropy, behind only BiCodec, indicating very efficient codebook usage.

\subsection{Codec inference efficiency}
We report time the encoding and decoding speed of all audio codecs in this work in Table \ref{tab:efficiency}. Although \modelname~contains more parameters than most baselines, it still achieves competitive inference speed, with the lowest encoding RTF and third-lowest total RTF, making it a strong choice for lightweight applications.

\begin{table}[h]
\centering
\resizebox{0.3\textwidth}{!}{%
\begin{tabular}{lc}
\toprule
Model & Entropy ($\uparrow$) \\
\midrule
BiCodec         & \B{0.995} \\
SpeechTokenizer & 0.984 \\
Kanade          & 0.976 \\
X-Codec2        & 0.965 \\
StableCodec     & 0.962 \\
FACodec         & 0.953 \\
DualCodec       & 0.923 \\
Mimi            & 0.914 \\
WavTokenizer    & 0.885 \\
PAST            & 0.824 \\
TiCodec         & 0.623 \\
\midrule
\modelname@12.5 & \U{0.985} \\
\bottomrule
\end{tabular}%
}
\caption{Normalized entropy of different audio codecs. Baseline results reported from \citet{DBLP:journals/corr/abs-2602-00594}.}
\label{tab:codebook}
\end{table}

\begin{table*}[h]
\centering
\resizebox{0.8\textwidth}{!}{%
\begin{tabular}{lc|ccc}
\toprule
Model & Params & Enc RTF (RTFx) & Dec RTF (RTFx) & Total RTF (RTFx)\\
& & $(\downarrow,\uparrow)$ & $(\downarrow,\uparrow)$ & $(\downarrow,\uparrow)$ \\
\midrule
Qwen3           & 171M  & 0.0035 (284x) & 0.0059 (169x) & 0.0094 (106x) \\
Mimi            & 96M   & 0.0032 (317x) & 0.0024 (415x) & 0.0055 (180x) \\
BiCodec         & 156M  & 0.0061 (164x) & 0.0024 (414x) & 0.0085 (117x) \\
X-Codec2        & 823M  & 0.0146 (68x)  & 0.0027 (368x) & 0.0174 (57x)  \\
WavTokenizer    & 81M   & \U{0.0026} (379x) & \U{0.0012} (802x) & \U{0.0039} (257x) \\
FocalCodec      & 145M  & 0.0027 (376x) & \B{0.0009} (1123x)& \B{0.0036} (282x) \\
Kanade          & 207M  & 0.0035 (284x) & 0.0021 (473x) & 0.0056 (177x) \\
\modelname@12.5 & 396M  & \B{0.0011} (928x) & 0.0034 (291x) & 0.0045 (221x) \\
\bottomrule
\end{tabular}%
}
\caption{Codec encoding/decoding efficiency. Total RTF measures the speed of both encoding and decoding.}
\label{tab:efficiency}
\end{table*}

\end{document}